\documentclass{elsart}

 \usepackage{graphicx}

\usepackage{amssymb}

\begin{document}

\begin{frontmatter}



\title{Two center shell model with Woods-Saxon potentials: adiabatic and
diabatic states in fusion}


\author[1,2]{A. Diaz-Torres}
\ead{diaz@th.physik.uni-frankfurt.de}
\author[2]{ and W. Scheid}
\address[1]{Institut f\"ur Theoretische Physik der
Johann Wolfgang Goethe--Universit\"at Frankfurt, Max von Laue Str. 1,
D--60438 Frankfurt am Main, Germany}
\address[2]{Institut f\"ur Theoretische Physik der
Justus--Liebig--Universit\"at Giessen, Heinrich--Buff--Ring 16,
D--35392 Giessen, Germany}

\begin{abstract}
A realistic two-center shell model for fusion is proposed, which is based on
two spherical Woods-Saxon potentials and the potential separable expansion method.
This model describes the single-particle motion in a fusing system.
A technique for calculating stationary diabatic states is suggested which makes
use of the formal
definition of those states, i.e., they minimize the radial nonadiabatic coupling between
the adiabatic states. As an example the system $^{16}$O + $^{40}$Ca $\to$ $^{56}$Ni
is discussed.
\end{abstract}

\end{frontmatter}


\section{Introduction}

The two-center shell model (TCSM) is a basic microscopic model to describe the single-particle
(sp) motion during a
heavy ion collision near the Coulomb barrier. The sp energies and wave functions are obtained
solving the Schr\"odinger equation with a phenomenological two-center mean field potential.
This approach was first introduced by the Frankfurt school \cite{Holzer,Maruhn}
and most of its applications so far in fission and fusion have been based on a double oscillator
potential \cite{Park}. Improved versions of TCSM based on oscillator potentials have been developed
for dealing with very asymmetric fission \cite{Mirea} and with asymmetric fragmentations
involving deformed fragments \cite{Radu}. The two-center problem for realistic finite depth potentials
has been solved with the wave function expansion method (usual diagonalization procedure)
in Refs. \cite{Pruess,Nuhn}. The authors of Ref. \cite{Pruess} made use of two spherical Woods-Saxon
potentials for describing the polarization of sp states in peripheral transfer reactions
like $^{40}$Ca + $^{16}$O and $^{208}$Pb + $^{12}$C, while in Ref. \cite{Nuhn} two deformed Gaussian
potentials were applied for the fusion of a deformed and arbitrarily orientated $^{13}$C projectile
on the $^{16}$O target.

In the present
work the two-center problem is solved using two Woods-Saxon (WS) potentials and the
potential separable expansion (PSE) method proposed by Revai \cite{Revai1,Revai2,Gareev,Milek}.
In contrast to Refs.\cite{Gareev,Milek}, where the PSE technique was employed within the
two-center framework to describe the sp motion in peripheral collisions
(the WS potential parameters were kept fixed) like $^{16}$O + $^{16}$O,
$^{17}$O + $^{13}$C and $^{17}$O + $^{16}$O, we now extend
the potential ansatz for large overlap between the nuclei by conserving the volume of the system
during the amalgamation process. For the sake of simplicity, we will use two spherical WS potentials,
whose parameters in a
realistic calculation have to be adjusted so as to fit the experimental sp spectra around
the Fermi level for the separated nuclei and for the spherical compound nucleus. For intermediate
internuclear distances these parameters can be interpolated as we will discuss below.
The present TCSM has the
following advantages compared to the traditional one based on two oscillator potentials
\cite{Maruhn}: (i) very
asymmetric reactions can be described as those used to form superheavy elements (e.g.,
$^{40}$Ca + $^{208}$Pb) without assumptions regarding the shape of the potential barrier
between the fragments, (ii) the relative level positions in the colliding nuclei are correct,
(iii) the two-center potential barrier between the nuclei is realistic,
(iv) the Coulomb interaction for protons can be explicitly included,
(v) the sp wave functions have a correct asymptotic behaviour \cite{Gareev},
and (vi) the continuum states like Gamow resonances can be included \cite{Milek} which
is relevant in reactions involving drip-line nuclei.
It is worth mentioning that the PSE method
has also been used in conjunction with both the one-center problem \cite{Ershov,Gyarmati}
and the two-center problem \cite{Fonseca} for nuclear structure studies.

In sect. 2, the PSE method to solve the adiabatic two-center problem is briefly presented
along with the
method to ensure the volume conservation of the fusing system. The adiabatic or
Born-Oppenheimer approximation is based on the idea that the relative motion of the nuclei is
much slower than the sp motion, so at each fixed internuclear distance the molecular
sp wave functions diagonalize the two-center sp hamiltonian. These states are called
$\textit{adiabatic}$ sp states and the nucleons occupy the lowest sp energy levels
in the two-center potential. The coupling between the adiabatic sp states,
called $\textit{nonadiabatic coupling}$, is induced by the kinetic energy operator related to
the relative motion of the nuclei. When the adiabatic sp states are well separated, this
coupling may be negligible and the adiabatic approximation is adequate. Whenever the
energy separation between two adiabatic sp states becomes small or vanishes, the coupling
may be large and the adiabatic approximation breaks down. In this case,
the sp state is a linear combination of adiabatic states. In such a situation a useful approach,
called stationary diabatic approximation, achieves the separation between the sp motion and
the relative motion of the nuclei. The stationary $\textit{diabatic}$ sp states minimize the
$\textit{radial}$ nonadiabatic coupling \cite{Delos}.
In the last part of sect. 2, we describe a technique to calculate the
stationary diabatic sp states following this formal definition.
The convenience of the diabatic sp states in the entrance phase of
heavy ion collisions around the Coulomb barrier was discussed by
Cassing and N\"orenberg in Ref. \cite{Cassing1}. In contrast to the adiabatic sp
motion, nucleons with a diabatic motion do not always occupy the lowest energy levels,
but remain in their diabatic levels during a collective motion of the system. The use of the
diabatic states accounts for the main part of the coherent coupling between collective and
intrinsic degrees of freedom.
Very recently in Ref. \cite{Alexis1} these states were applied for the modelling of the
compound nucleus formation in the fusion of heavy nuclei.
In sect. 3 numerical results for $^{16}$O + $^{40}$Ca $\to$ $^{56}$Ni are discussed,
while conclusions are drawn in sect. 4.

\section{The TCSM}

We will follow the method proposed by Gareev et al. in Ref. \cite{Gareev} and also used by
Milek and Reif in Ref. \cite{Milek}, which is based on the $\textit{expansion of potentials}$
in terms of, e.g., harmonic oscillator
functions (PSE method), i.e, $V \approx \sum_{ij}^{N} |i> V_{ij}<j| \equiv V_{sep}$.
The accuracy of
the results obtained depends only on the accuracy of the expansion of the WS potentials.
The Schr\"odinger equation with approximate potentials $V_{sep}$ is solved exactly.

The finite depth nuclear potential belongig to each fragment ($s=1,2$) is chosen to be the
following spherical WS with a spin-orbit term

\begin{equation}
V_s(r)=-V_{0s}f^s(r) + \frac{1}{2}\lambda_s (\frac{\hbar}{m_0c})^2 V_{0s} \frac{1}{r}\frac{df_{so}^s}{dr}
(\textbf{l} \cdot \textbf{s}), \label{eq_17}
\end{equation}
where $\hbar /m_0c = 0.21$ fm, $f$ and $f_{so}$ are the same function, but with different parameters, i.e.,
$f_{(so)}^s (r)=\{1+\exp[(r-R_{0(so)}^s)/a_{0(so)}^s]\}^{-1}$.
For protons, the Coulomb potential
$V_{Coul}^s$ is taken to be that of a uniformly charged sphere with charge
$Z_s e$ ($Z_s$ being the
total charge of each fragment) and  the radius $R_{c}^s$, which is added to
expression (\ref{eq_17}). The potentials (\ref{eq_17}) are placed at the position $\bf{R}_s$
in the center of mass system and thus the two-center potential reads as

\begin{equation}
V = e^{-i \textbf{R}_1 \hat{k}}\ V_1\ e^{i \textbf{R}_1 \hat{k}} +
e^{-i \textbf{R}_2 \hat{k}}\ V_2\ e^{i \textbf{R}_2 \hat{k}},
\label{eq_18}
\end{equation}
where $\hat{k}=\hbar^{-1}\hat{p}$ is the sp wave-number operator.
Each potential (\ref{eq_17}) is then represented approximately ($V_{sep}^s$)
within a truncated sp harmonic oscillator basis,
$\{|\nu >, \nu = 1, \ldots N \}$,
with the spin-angular part coupled to the total angular momentum $j$ with projection $m$,
e.g., in the momentum representation

\begin{equation}
|\nu> = |nljm> = g_{nl}(k)\cdot [i^{-l}Y_l (\hat{\bf{k}})
\otimes \chi_{\frac{1}{2}}(s) ]^{j}_{m}. \label{wf_basis}
\end{equation}

The harmonic oscillator basis has the advantage that all matrix elements needed
can be calculated analytically (e.g., see Appendix A in Ref. \cite{Milek}) and
this basis adopts
essentially the same mathematical form in momentum and coordinate representations.
In the coordinate representation (spherical coordinates), the radial part of the harmonic oscillator
wave function can be written as $<r|nl>=h_{osc}^{-3/2} \phi_{nl} (x)$, where $h_{osc}$ is the
oscillator length and the adimensional $x=h_{osc}^{-1}\ r$.
In the momentum representation,
$<k|nl> \equiv g_{nl}(k) = (-1)^n\ i^l\ h_{osc}^{3/2}\ \phi_{nl} (\xi)$ being the adimensional
$\xi=h_{osc}\ k$. The phase factor $i^{-l}$ in the definition of the basis set (\ref{wf_basis})
removes the imaginarity of $g_{nl}(k)$.
The function $\phi_{nl}$ is the reduced oscillator function, i.e.,

\begin{equation}
\phi_{nl} (\xi)=\sqrt{\frac{2n!}{\Gamma (n+l+1/2)}}\cdot \xi ^{l}
\cdot e^{-\xi^{2}/2}\cdot L_{n}^{l+1/2}(\xi^{2}), \label{radial_basis}
\end{equation}
where $L_{n}^{l+1/2}(\xi^{2})$ is a Laguerre polynomial. According to
Ref. \cite{Ring}, the oscillator lenght $h_{osc}^{s}$
associated with each basis set can be calculated with the expression
$h_{osc}^{s}=0.84\cdot R_{0}^s\cdot A_{s}^{-1/6}$,
being $A_{s}$ the mass number of each fragment.

\textit{Volume conservation}. To describe the amalgamation process of the two nuclei, the
potential parameters have to be interpolated between their values for the separated nuclei and
the spherical compound nucleus. The parameters can be correlated by assuming the condition that
the volume enclosed by certain equipotential surface $V_0$ of the two-center potential $V$ is
conserved for all values of $R$, i.e.,

\begin{equation}
\int dv\ \Theta (V_0 -V)=v_0, \label{eq_23}
\end{equation}
where $\Theta$ is the Heaviside step function, $V_0$ the fixed equipotential surface corresponding
to the Fermi level of the fused system,
and $v_0$ the volume enclosed by the equipotential surface $V_0$. As the two-center
potential $V$ in (\ref{eq_23}), we use $\widetilde{V_1} + \widetilde{V_2}$ where
$\widetilde{V_s}$ refers to the first term of (\ref{eq_17}), i.e.,
the spin-orbit part is neglected.
For the sake of simplicity, we will assume that the nuclear shape is the same for
neutrons and protons.

The parameters $\mathcal{P} = \{V_{0s}, R_{0(so)}^s, a_{0(so)}^s, \lambda_s, R_{c}^s,
Z_s\}$ are interpolated by means of the unknown function $y(R)$ \cite{Nuhn}

\begin{equation}
\mathcal{P}(R) = \mathcal{P}(R=0)\ [1-y(R)] + \mathcal{P}(R \rightarrow \infty)\ y(R), \label{eq_23b}
\end{equation}
where $y(R)$ satisfies the boundary conditions $y(R=0) = 0$ and $y(R\rightarrow \infty)= 1$. The
function $y(R)\in [0,1]$ is numerically calculated by inserting (\ref{eq_23b}) into (\ref{eq_23}) and
requiring the conservation of the volume. The parameters $\mathcal{P}(R \rightarrow \infty)$ are
unambiguously
found fitting the experimental sp spectra around the Fermi level for the separated nuclei, but this
is not the case for the individual fragment potentials at $R=0$ where only the sp spectra of
the compound system are known. At $R=0$ we adopt that $R_{0(so)}^s, a_{0(so)}^s, \lambda_s$ and
$R_{c}^s$ are the same as those
for the compound nucleus, but $V_{01} + V_{02} = V_{0}^{CN}$ and
$Z_{1} + Z_{2} = Z_{CN}$. A new parameter $\alpha = V_{01} / V_{0}^{CN} =
Z_{1}/ Z_{CN}$ is introduced and it determines the values of $V_{0s}$ and $Z_{s}$
at $R=0$. This $\alpha \in (0,1)$ parameter is chosen in such a way that the splitting of the sp levels
for small $R$ is similar to that in the Nilsson model.

$\textit{Adiabatic states}$. In the present method the two-center problem is solved
in the momentum representation. The approximate potential $V_{sep}^s$ reads as

\begin{equation}
V_{sep}^s=\sum_{\nu, \mu = 1}^{N} |s\nu> V_{\nu \mu}^s <s\mu |, \label{eq_18b}
\end{equation}
where $\nu,\mu = \{ nljm\}$. The number $N$ of basis states included in the expansion
(\ref{eq_18b}) is defined by $l_{max}$ (number of partial waves in which the potential acts) and
$n_{max}$ (the number of separable terms in each partial wave). The values of $l_{max}$ and
$n_{max}$ are determined by the convergence of the sp energies which is accelerated
using the technique of the Lanczos $\sigma$-factors \cite{Alexis3}. For bound states the formal
solution of the Schr\"odinger equation is as follows

\begin{equation}
|\varphi> = G_0(E)V |\varphi>,
\quad G_0(E)= (E - \frac{\hbar^2\hat{k}^2}{2m_0})^{-1}, \label{eq_19}
\end{equation}
where $G_0$ is the Green operator for the free sp motion.
Inserting (\ref{eq_18}) and (\ref{eq_18b}) into (\ref{eq_19}), the expression (\ref{eq_19})
reads as

\begin{equation}
|\varphi> = G_0(E)\sum_{\nu, \mu = 1}^{N} \sum_{s=1}^{2} V_{\nu \mu}^s\ A_{s\mu}\
e^{-i \textbf{R}_s \hat{k}}\ |s\nu>, \label{eq_19b}
\end{equation}
where

\begin{equation}
A_{s\mu}=<s\mu|e^{i \textbf{R}_s \hat{k}}|\varphi>. \label{eq_19c}
\end{equation}

The coefficients $A_{s\mu}$ are the amplitude of the molecular sp wave function $|\varphi>$
with respect to the moving basis states located at $\textbf{R}_s$.
Multiplying (\ref{eq_19b}) from
the left by $<s\mu|e^{i \textbf{R}_s \hat{k}}$ and with (\ref{eq_19c}),
the following set of linear
algebraic equations for the coefficients $A_{s\mu}$ is obtained:

\begin{equation}
\sum_{\mu'=1}^{N}\sum_{s'=1}^{2}[\delta_{ss'} \delta_{\mu \mu'} -
\sum_{\nu=1}^{N} <s\mu|G_0(E)e^{i (\textbf{R}_s - \textbf{R}_{s'}) \hat{k}}|s'\nu>\
V_{\nu \mu'}^{s'}]\ A_{s'\mu'}\ = 0. \label{eq_20}
\end{equation}

The solvability condition of the system (\ref{eq_20}) is that its determinant vanishes.
This determines the energy eigenvalues $E$ which appear as a parameter in the matrix elements
containing $G_0$. With the eigenvalues $E$, the eigenvectors $|\varphi>$ are obtained solving
the system (\ref{eq_20}) for the coefficients $A_{s\mu}$ and requiring the normalization of the
state vectors $|\varphi>$. These states are the adiabatic sp states.

The quantum numbers denoting the molecular sp states arise from the symmetry properties of the
two-center problem. Since the potential (\ref{eq_18}) is invariant around the axis connecting
the centers of both potentials ($\textbf{R}=\textbf{R}_1 - \textbf{R}_2$),
the projection of the total angular momentum $j$ along this
axis denoted by $\Omega$ is conserved. This axis is chosen as the quantization axis and the
sp levels are labelled by $\Omega$ and an additional index $indx$ that distinguishes different states
with the same $\Omega$. The index $indx$ could refer either to the asymptotically good quantum numbers
($\tilde{n}lj$) of the
state to which the level goes for large $R$ or just to an index $\tilde{n}$ counting the levels.
In the intrinsic (or rotating) molecular reference frame
($\textbf{R}_1=\textbf{R}$ and $\textbf{R}_2=0$) the matrix
elements containing $G_0$ are as follows ($\hat{j}=(2j+1)^{1/2}$)

\begin{equation}
s=s':\quad <nlj\Omega|G_0(E)|n'l'j'\Omega'>=\delta_{ll'}\delta_{jj'}\delta_{\Omega \Omega'}\
<nl|G_0(E)|n'l'>, \label{eq_21a}
\end{equation}

\begin{eqnarray}
s\neq s':&&\nonumber \\
&&<nlj\Omega|G_0(E)e^{i \textbf{R}_{ss'} \hat{k}}|n'l'j'\Omega'>=
\delta_{\Omega \Omega'}(-1)^{\Omega+\frac{1}{2}} \cdot
\sum_{L}i^{L+l-l'}\nonumber\\
&&Q_{ss'}\ \hat{l}\hat{j}\hat{l'}\hat{j'}\ (l0l'0|L0)
(j\Omega j' -\Omega|L0)\
\Bigg\{ \begin{array}{ccc}
L & l' & l \\
\frac{1}{2} & j & j'
\end{array} \Bigg\}\nonumber \\
&& <nl|G_0(E)\ j_L(kR)|n'l'>, \label{eq_21b}
\end{eqnarray}
where

\begin{equation}
Q_{ss'}=\Bigg\{ \begin{array}{c}
1,\  ss'=12 \\
(-1)^L,\ ss'=21
\end{array}, \label{eq_22a}
\end{equation}
\begin{equation}
<nl|G_0(E)|n'l'>=-\frac{2m_0}{\hbar^2}\int_{0}^{\infty}\frac{g_{nl}^{*}(k)g_{n'l'}(k)}
{k^2 + \gamma^2}\ k^2dk,\quad \gamma^2=-\frac{2m_0 E}{\hbar^2}, \label{eq_22b}
\end{equation}
\begin{equation}
<nl|G_0(E)\ j_L(kR)|n'l'>=-\frac{2m_0}{\hbar^2}\int_{0}^{\infty}
\frac{g_{nl}^{*}(k)g_{n'l'}(k)j_L(kR)}
{k^2 + \gamma^2}\ k^2dk. \label{eq_22c}
\end{equation}

In (\ref{eq_22c}) $j_L$ refers to the spherical Bessel function. The matrix elements
(\ref{eq_21b}) are obtained using (i) the expansion of
$e^{i \textbf{R}_{ss'} \textbf{k}}$ in terms of the spherical functions

\begin{equation}
e^{i \textbf{R}_{ss'} \textbf{k}}=4\pi \sum_{L=0}^{\infty}\sum_{M=-L}^{L} i^L
j_L(kR_{ss'})Y_{LM}^{*}(\hat{\bf{R}}_{ss'})Y_{LM}(\hat{\bf{k}}), \label{expansion}
\end{equation}

and (ii) general properties of the angular momentum algebra \cite{Varshalovich}.

The sp states $|indx\ \Omega>$ calculated with
(\ref{eq_20})-(\ref{eq_22c}) refer to adiabatic molecular sp states in the
intrinsic (or rotating) reference frame, where the amplitude of the molecular wave functions
solving (\ref{eq_20}) are denoted by $B_{s\beta\Omega}$ with $\beta = \{nlj\}$.
Please note that these amplitudes depend
only on the relative distance $R$. They are independent of the orientation angles $\hat{R}$
of the molecular axis, since these are no longer contained in the matrix elements
(\ref{eq_21a})-(\ref{eq_22c}). The rotation of the molecular axis
with respect to a space-fixed axis is then included in (\ref{eq_19b}) by means of the Wigner
$D^{j}_{m \Omega}(\hat{R})$ rotation matrices, so the $|indx\ \Omega>$ adiabatic molecular sp
states in the laboratory reference frame read as follows

\begin{eqnarray}
|indx\ \Omega> = G_0(E_{indx,\Omega})\sum_{\beta m, \beta'}\sum_{s=1}^{2}
V_{\beta m \beta' \Omega}^s\ B_{s\beta' \Omega}\ D^{j}_{m \Omega}(\hat{R})\
e^{-i \textbf{R}_s \hat{k}}\ |s\beta m >. \nonumber \\ \label{eq_22e}
\end{eqnarray}

$\textit{Stationary diabatic states}$. A natural way of introducing the
stationary diabatic sp states is to impose the condition that the radial coupling
operator $(-i\hbar\partial /\partial R + \mathcal{A})$ be zero or negligibly small
in these states \cite{Delos}, where $\mathcal{A}$ identifies and cancels that
spurious portion
of the coupling induced by $-i\hbar\partial /\partial R$ that represents merely the
translation of basis functions along with moving nuclei.
Fortunately only inspection is needed to remove from the $-i\hbar\partial /\partial R$ matrix
\textit{that part which corresponds simply to displacement}, as we will show next.
Then what remains after such background correction is the actual nonadiabatic coupling which
vanishes for large internuclear distances. The diabatic representation
is obtained by diagonalization of this coupling in a truncated set
of the adiabatic sp states (\ref{eq_22e}) with the same quantum number $\Omega$, i.e.,
$<indx\ \Omega|(-i\hbar \partial /\partial R + \mathcal{A})|indx'\ \Omega>$.
Only couplings between neighbouring levels ($indx'=indx \pm 1$) are included.
In defining the matrix of the radial nonadiabatic couplings that is
diagonalized, we neglect the couplings whose strength is smaller than a threshold
value $\gamma_{thr}$. The value of $\gamma_{thr}$ used in the calculations
should lead to the isolation of the strongest radial nonadiabatic couplings.
Strong radial nonadiabatic couplings are expected to occur at the
avoided crossing of the adiabatic molecular sp levels.

The adiabatic molecular sp states (\ref{eq_22e}) can be re-written as

\begin{equation}
|indx\ \Omega> = G_0(E_{indx,\Omega}) \{ \sum_{\nu = 1}^{N_1}
C_{1\nu }^{\Omega} e^{-i \textbf{R} \hat{k}}\ |1\nu> + \sum_{\nu = 1}^{N_2}
C_{2\nu}^{\Omega} |2\nu>\}, \label{24a}
\end{equation}
where the coefficients $C_{s\nu}^{\Omega}=\sum_{\beta'} V_{\nu \beta'\Omega}^s B_{s\beta'\Omega}$.
Here, we identify $\nu = \{\beta m\}$ and the Wigner rotation matrices are now
included in the basis $|s\nu>$. The radial derivative
$\partial /\partial R$ of (\ref{24a}) is as follows

\begin{eqnarray}
\frac{\partial}{\partial R}|indx\ \Omega>=&&\frac{\partial G_0(E_{indx,\Omega})}{\partial R}
\{ \sum_{\nu = 1}^{N_1}
C_{1\nu }^{\Omega} e^{-i \textbf{R} \hat{k}}\ |1\nu> +
\sum_{\nu = 1}^{N_2} C_{2\nu}^{\Omega} |2\nu>\} \nonumber \\
&&+\ G_0(E_{indx,\Omega}) \{ \sum_{\nu = 1}^{N_1}
\frac{\partial C_{1\nu }^{\Omega}}{\partial R} e^{-i \textbf{R} \hat{k}}\ |1\nu> +
\sum_{\nu = 1}^{N_2}\frac{\partial C_{2\nu}^{\Omega}}{\partial R}\ |2\nu>\} \nonumber \\
&&+\ G_0(E_{indx,\Omega}) \sum_{\nu = 1}^{N_1}C_{1\nu}^{\Omega}
\frac{\partial e^{-i \textbf{R} \hat{k}}}{\partial R}\ |1\nu>. \label{24b}
\end{eqnarray}

The last term of (\ref{24b}) is responsible for the unphysical effect (i.e. non vanishing
asymptotic couplings owing to the
displacement of the moving basis) on the radial nonadiabatic transitions discussed above and
will therefore be removed.
Consequently, from expression (\ref{24b}) one can observe that the radial nonadiabatic coupling
vanishes for $R\rightarrow\infty$ because
$\partial G_0(E_{indx,\Omega})/\partial R=-\partial E_{indx,\Omega}/\partial
R\cdot G_0^2(E_{indx,\Omega})\rightarrow 0$ and
$\partial C_{s\nu}^{\Omega}/\partial R \rightarrow 0$
when the molecular sp states approach the fixed sp states of the separated nuclei.
In Appendix \ref{Radial_matrix_elements}
we give expressions for the calculation of the actual
radial nonadiabatic coupling matrix elements including the first four terms of (\ref{24b}) only.

After diagonalizing the actual radial nonadiabatic coupling we obtain the unitary transformation
matrix $S_{ij}^{\Omega}$ between the adiabatic and diabatic representations, where $i,j$ refer to
the values of $indx$. The diabatic sp wave functions and energy levels are obtained as
follows

\begin{equation}
|i\ \Omega>_{diab} = \sum_{j} s_{ij}^{\Omega}\ |j\ \Omega>_{adiab}, \label{26a}
\end{equation}

\begin{equation}
E_{i\Omega}^{diab} = \sum_{j} |s_{ij}^{\Omega}|^2 E_{j\Omega}^{adiab}. \label{26b}
\end{equation}

For a given $\Omega$ the adiabatic sp levels $E_{i\Omega}^{adiab}$ as a function of $R$
(correlation diagram \cite{Nikitin}) show avoided crossings because of the noncrossing rule
by von Neumann and Wigner \cite{Neumann}, while the diabatic levels $E_{i\Omega}^{diab}$
reveal real crossings. The diabatic states (\ref{26a})-(\ref{26b})
can cross each other because they are not solutions of an eigenvalue problem.

There are approximating methods \cite{Nikitin} for calculating diabatic states, namely the
maximum symmetry method and the maximum overlap method. These techniques were applied in
Ref. \cite{Lukasiak1} to obtain diabatic states with a TCSM based on harmonic oscillators.
In Refs. \cite{Alexis1,Alexis2} we made use of the maximum symmetry method to calculate diabatic
potential energy surfaces of near symmetric reactions.
Other methods have been developed in quantum chemistry for the calculation of such a
diabatic basis, e.g., see Ref. \cite{Diabbasis}.

\section{Numerical results}

As an example of the realistic TCSM discussed above we present calculations for neutrons
and protons in the reaction
$^{16}$O + $^{40}$Ca $\to$ $^{56}$Ni. The potential parameters for the separated nuclei and
the compound nucleus as well as the maximal number of nodes $n_{max}$ and partial waves
$l_{max}$ included in the harmonic oscillator basis for converged values of the sp
energies are given in Table 1. These parameters approximately reproduce the
experimental sp levels around the Fermi energy, which was shown in Refs. \cite{Pruess,Soloviev}.

\begin{table}
\caption{Potential parameters for the separated nuclei and the
compound nucleus: The potential depth is
given in MeV, the radii and diffusenesses in fm, while the strength $\lambda$ of the spin-orbit
potential is adimensional. The radii $R_{0(so)}= r_{0(so)}A^{1/3}$ being A the mass number of
the nucleus.
See text for further details.}
\vspace{1cm}
\begin{tabular}{cccccccccc}
\hline
\multicolumn{10}{c}{Neutrons} \\
Nucleus& $V_0$ & $r_0$ & $a_0$ & $\lambda$ & $r_{so}$ & $a_{so}$ & $R_c$ & $n_{max}$
&$l_{max}$ \\
\hline
$^{16}$O$^{a}$ & 74.90 & 0.97 & 0.60 & 19.50 & 1.10 & 0.55 &  & 4 & 3 \\
$^{40}$Ca$^{a}$ & 51.40 & 1.28 & 0.76 & 24 & 1.37 & 0.55 &  & 4 & 3 \\
$^{56}$Ni$^{b}$ & 46.20 & 1.31 & 0.62 & 18.73 & 1.31 & 0.62 &  & 4 & 3 \\
\multicolumn{10}{c}{Protons} \\
$^{16}$O$^{a}$ & 79.10 & 0.95 & 0.59  & 19.15 & 1.10 & 0.55 & 2.77 & 4 & 3 \\
$^{40}$Ca$^{a}$ & 70.80 & 1.05 & 0.67 & 19.22 & 1.05 & 0.55 & 3.83 & 4 & 3 \\
$^{56}$Ni$^{b}$ & 53.70 & 1.24 & 0.63 & 13.97 & 1.31 & 0.63 & 4.74 & 4 & 3 \\
\hline
\end{tabular}
\\$^{a}$From Ref. \cite{Pruess}, $^{b}$From Ref. \cite{Soloviev}
\end{table}

\subsection{Neutrons}

Fig. 1 shows the function $y(R)$ derived from the condition of volume conservation
for different values of the $\alpha$ parameter which determines both the partition of the
compound nucleus potential depth into two parts belonging to each fragment and their
charges. The numerically calculated function $y(R)$ is used in expression (\ref{eq_23b})
to interpolate the two-center potential parameters. It is observed that (i)
$y(R)$ decreases from 1 around the contact configuration ($R \approx 7.3$ fm) up to 0 for the
spherical compound nucleus ($R=0$ fm), and (ii) $y(R)$ very strongly depends on
the $\alpha$ parameter at small radii R, while this is not the case around the contact point.
The strong effect of the $\alpha$ parameter on the neutron sp spectra at small distances $R$
can be seen in Fig. 2.

\begin{figure}
\begin{center}
\includegraphics[width=15.0cm]{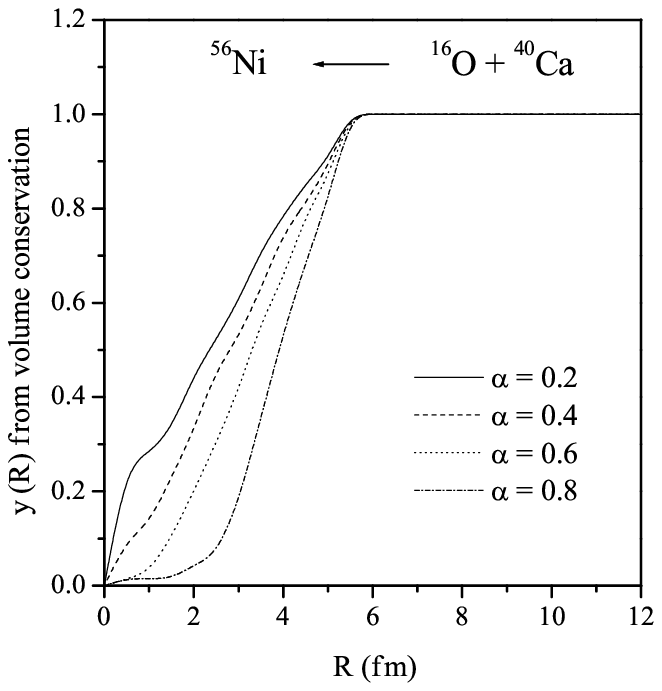}
\end{center}
\caption{The function $y(R)$ derived from the condition of volume conservation for different
values of the $\alpha$ parameter. This is used to
interpolate the neutron and proton two-center potential parameters as a function of the
radius $R$ between the nuclei. See text for further details.}
\end{figure}

In Fig. 2 we show normalized neutron sp levels with $\Omega = 1/2$ as a function of the
quadrupole deformation parameter
$\beta$ of the $^{56}$Ni compound nucleus for two values of the $\alpha$ parameter, namely
$\alpha=0.4$ (right panel) and $\alpha=0.8$ (middle panel). Moreover, these level diagrams
are compared to the levels of the Nilsson model \cite{Irving} (left panel).
The normalization is given by the first Nilsson-level and the $\beta$ value is
obtained from the quadrupole moment of the nuclear shape.
From this figure it is concluded that a realistic value of $\alpha$ is about 0.8 because,
as expected, the behaviour of the two-center sp levels is similar to that in the Nilsson model.
In the following we will use $\alpha = 0.8$.

\begin{figure}
\begin{center}
\includegraphics[width=15.0cm]{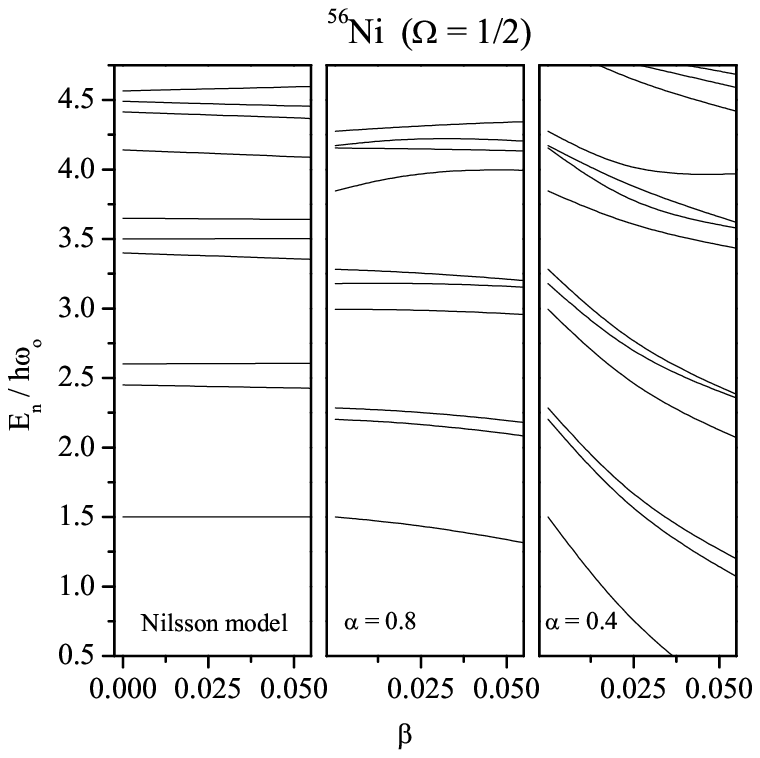}
\end{center}
\caption{Normalized neutron levels with $\Omega = 1/2$ as a function of the quadrupole deformation
$\beta$ of the compound nucleus $^{56}$Ni for different values of the $\alpha$ parameter:
$\alpha=0.4$ (right panel) and $\alpha=0.8$ (middle panel). The left panel shows the levels
of the Nilsson model. See text for further details.}
\end{figure}

Fig. 3 shows the central part of the neutron two-center potential along the $z$-axis
for fixed values of the distance $R$ between the fragments. The nuclear shape associated
with each potential is also included in this figure. Near the
contact configuration ($R = 7$ fm) the interfragment potential barrier
practically disappears and, therefore, most of the nucleons may move in the
whole volume of the system. From distances $R \lesssim 4$ fm the two-center potential
essentially adopts a one-center character. It is worth mentioning that the
two-center potential barrier is naturally formed in the present model by the superposition
of the tail of the two WS potentials.

\begin{figure}
\begin{center}
\includegraphics[width=15.0cm]{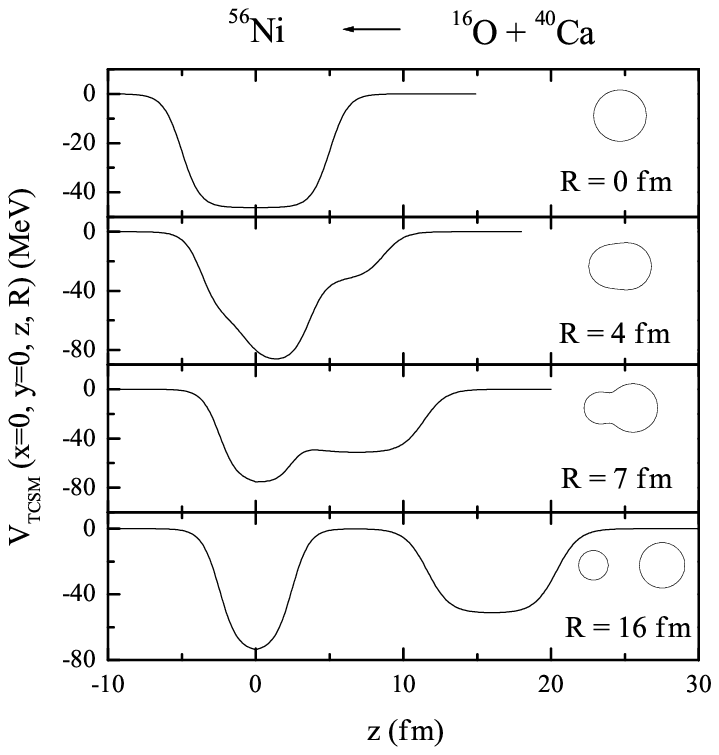}
\end{center}
\caption{The central part of the neutron two-center potential along the z-axis for fixed
radii $R$ between the nuclei. The nuclear shape associated with each potential is also
included. See text for further details.}
\end{figure}

Fig. 4 shows the adiabatic (left panel) and diabatic (right panel) neutron levels with
$\Omega = 1/2$ as a function of the internuclear distance $R$. For small and large radii
we can observe the shell structure of the compound nucleus and the separated fragments,
respectively. For distances 3 fm $\lesssim R \lesssim $ 7 fm
sp levels are strongly polarized and the shell structure of the separated fragments is
practically dissolved. Here sp excitations and transfer processes essentially occur around
the avoided crossing of the molecular adiabatic sp levels due to the effect of the radial
nonadiabatic couplings on the sp motion. Most of the avoided crossings in the adiabatic level
diagram turn into real crossings in the diabatic level diagram. This happens where the
radial nonadiabatic coupling between the adiabatic sp levels shows huge peaks
which can be seen in Fig. 5.

\begin{figure}
\begin{center}
\includegraphics[width=15.0cm]{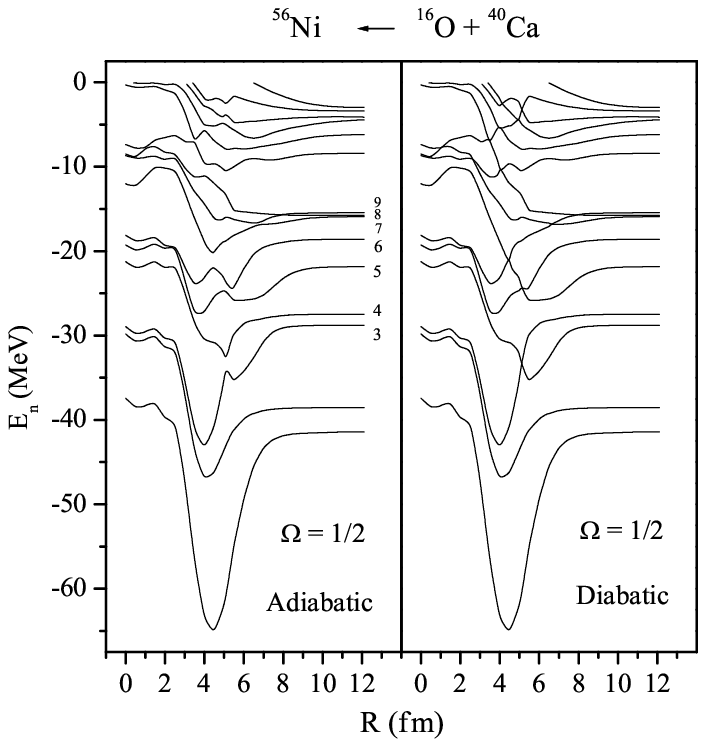}
\end{center}
\caption{Adiabatic (left panel) and diabatic (right panel) neutron levels with
$\Omega=1/2$ as a function of the radius $R$ between the nuclei. Some adiabatic levels
are labeled by 3-9. See text for further details.}
\end{figure}

Fig. 5 shows the absolute value of the radial nonadiabatic couplings between some
neighbouring adiabatic levels of Fig. 4 (left panel) as a function of the
internuclear radius: levels 3-4 (solid curve), levels 5-6 (dotted curve) and
levels 8-9 (dashed curve). For large distances the couplings vanish as expected.
The more bound the levels are, the smaller is the radius $R$ in which the coupling
vanishes, e.g., comparing levels 3-4 (solid curve) and levels 8-9 (dashed curve).
The huge peaks localize the position of an avoided crossing where a nucleon
transition (Landau-Zener effect) may occur with a large probability. These peaks are the
key to calculating the diabatic states which are obtained by minimizing those strong
couplings only. The threshold value for the strength of the couplings
was $\gamma_{thr}=1.4$. It is important to stress that
strong transitions between molecular sp levels as that between the levels 8-9
(dashed curve) at $R \approx 8$ fm can also occur for distances larger than that in the
contact configuration.
In these situations Landau-Zener nucleon transitions may be reflected in the
excitation function of direct reaction processes \cite{Park}.

\begin{figure}
\begin{center}
\includegraphics[width=15.0cm]{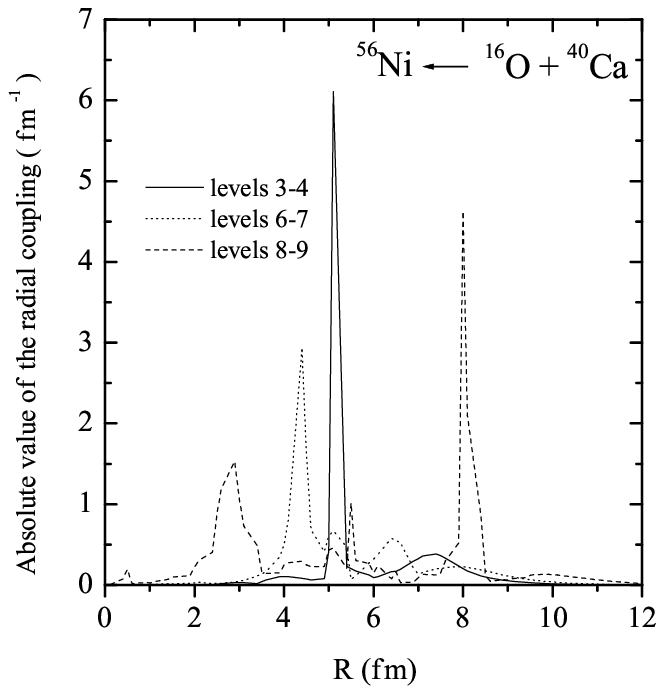}
\end{center}
\caption{Absolute value of the radial nonadiabatic coupling between some neighbouring
adiabatic levels of Fig. 4 (left panel) as a function of the radius $R$ between the nuclei.
See text for further details.}
\end{figure}

In Fig. 6, the adiabatic neutron levels correlation diagram including all bound states with
different $\Omega$ values is presented, i.e., $\Omega=1/2$ (solid curves), $\Omega=3/2$
(dashed curves),
$\Omega=5/2$ (dotted curves) and $\Omega=7/2$ (dashed-dotted-dotted curve).
The states with the same $\Omega$ value are coupled by the radial coupling discussed above,
while those states with $\Omega$ values differing by one unit
in $\Omega$ (e.g., $\Omega=1/2$ and $\Omega=3/2$) are coupled in the laboratory reference
frame by the so-called rotational (or Coriolis) coupling which is maximal at the real crossing
of these states \cite{Park}.

\begin{figure}
\begin{center}
\includegraphics[width=15.0cm]{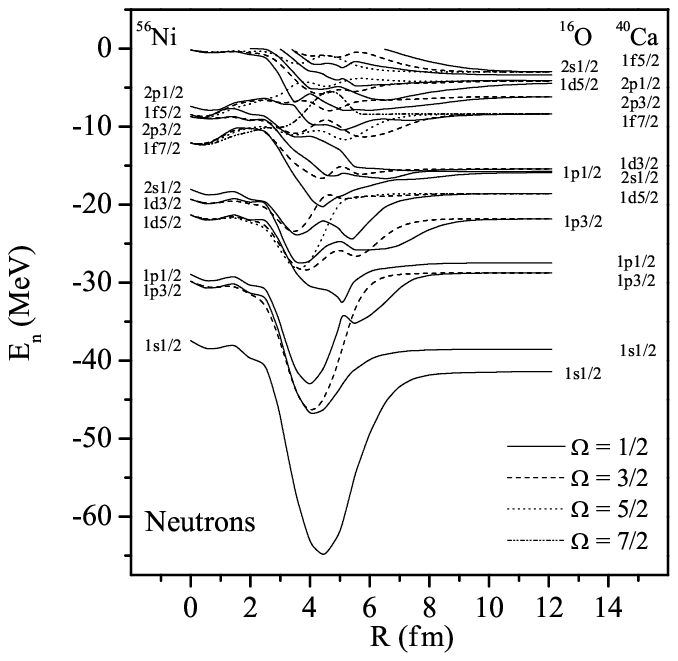}
\end{center}
\caption{}
\caption{The adiabatic neutron levels correlation diagram including all
bound states with different $\Omega$ values. See text for further details.}
\end{figure}

Both the adiabatic and the diabatic sp states with different $\Omega$ values,
as mentioned above, are coupled in the laboratory system by the rotational coupling.
The rotational coupling
($\sim \bf{j} \cdot \bf{I}$, where $\bf{j}$ is the nucleon total angular momentum and
$\bf{I}$ is the total angular momentum of the system) appears
in a non central collision due to the rotation of the molecular internuclear axis with respect to
a space-fixed axis. Please note that the sp wave-functions ((\ref{eq_22e}) and (\ref{26a}))
depend on the Wigner rotation matrices. One can expect that the rotational coupling
(which increases with decreasing internuclear radii $R$, i.e., $\sim 1/R^2$)
is important in light ion collisions owing
to the small moment of inertia of the system, but it is not the case in a heavy ion reaction.
In fusion of light systems (where the fusion process is already determined by the penetration of
the external Coulomb barrier, i.e., small overlap between the nuclei) the relevant radial region
for the rotational coupling is governed by the usually strongly absorptive nucleus-nucleus
potential. Terlecki et al. \cite{Terlecki} have found in the reaction
$^{13}$C+$^{13}$C that the effect of the rotational coupling
on the inelastic excitation cross sections is much smaller than the effect of the radial
coupling. We would like to add that Imanishi and von Oertzen
have introduced in Ref. \cite{Imanishi} the so-called Rotating Molecular Orbitals (RMO) in
the standard coupled reaction channels formalism in order to avoid numerical problems with the
rotational coupling at small internuclear radii $R$. To our knowledge, there has been no systematic
study of the effect of rotational couplings on fusion and scattering cross sections. Further work
on this issue is required.

\subsection{Protons}

Since we have assumed that the nuclear shape
is the same for neutrons and protons, the same function $y(R)$ of Fig. 1
($\alpha=0.8$, dashed-dotted curve) is applied
in interpolating the proton two-center potential parameters.
In Fig. 7 like in Fig. 3, the central part of the proton two-center potential along the
$z$-axis is shown for different fixed internuclear distances $R$. Apart from (i) the positive
tails of the two-center potential for large $z$ values due to the Coulomb interaction and
(ii) a higher barrier between the fragments at $R = 7$ fm,
the change of the two-center potential with $R$ is similar to that in Fig. 3 for neutrons.
Fig. 8 shows the adiabatic (left panel) and diabatic (right panel) correlation diagram
of the proton levels with $\Omega=1/2$. The same features as those
observed for neutrons in Fig. 4 can be seen here.
In order to isolate the strong peaks of the proton radial nonadiabatic coupling,
a threshold value $\gamma_{thr}=1.7$ for the strength of the couplings was needed
in this case to calculate the diabatic levels. Fig. 9 is like Fig. 6, but for protons.
The level diagram shows similar features as those in Fig. 6, only the levels are shifted
up due to the Coulomb interaction.

\begin{figure}
\begin{center}
\includegraphics[width=15.0cm]{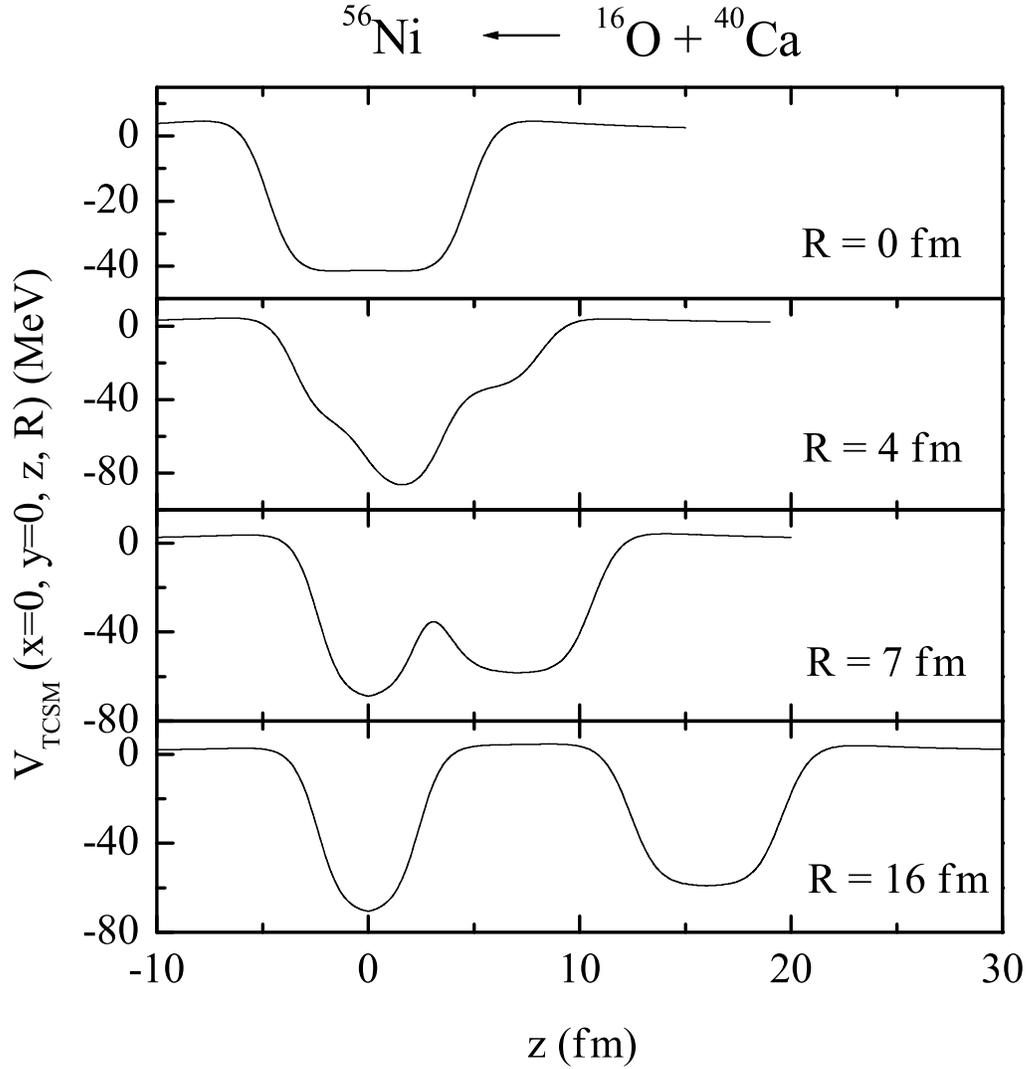}
\end{center}
\caption{The same as Fig. 3, but for protons. See text for further details.}
\end{figure}

\begin{figure}
\begin{center}
\includegraphics[width=15.0cm]{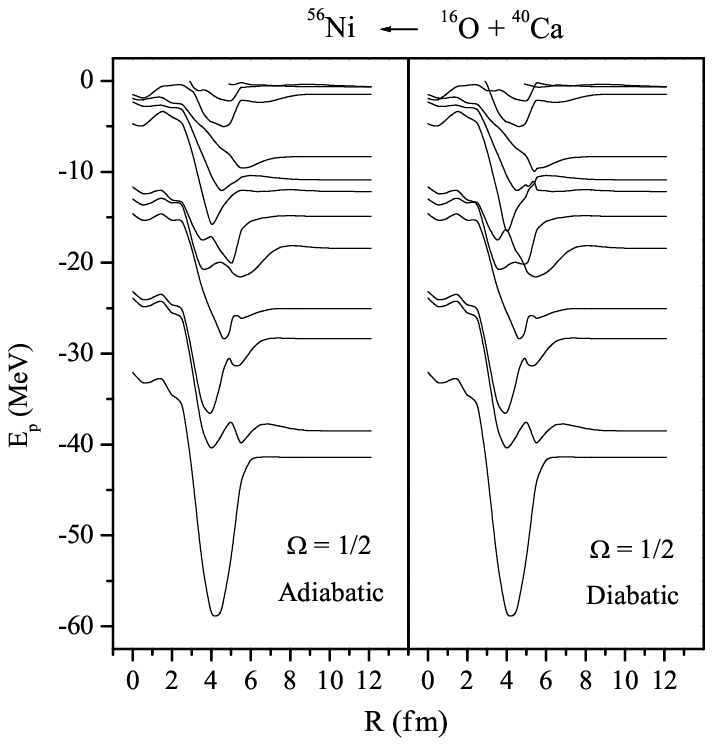}
\end{center}
\caption{The same as Fig. 4, but for protons. See text for further details.}
\end{figure}

\begin{figure}
\begin{center}
\includegraphics[width=15.0cm]{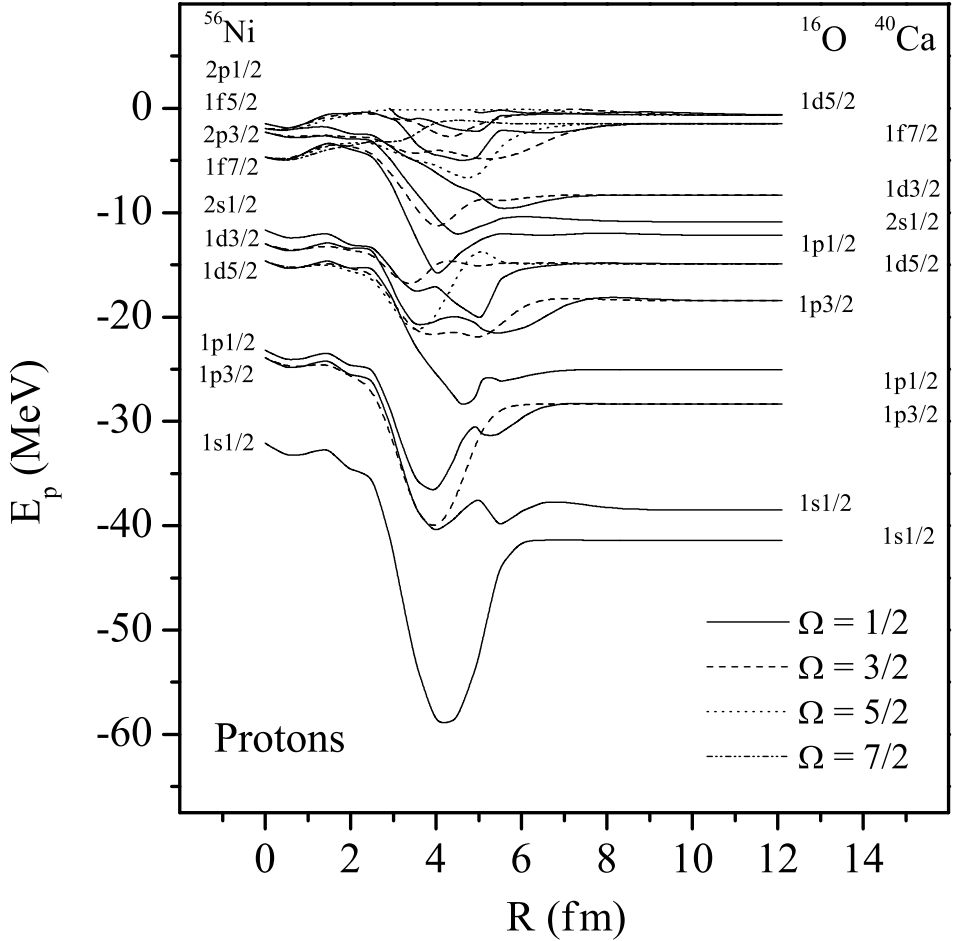}
\end{center}
\caption{The same as Fig. 6, but for protons. See text for further details.}
\end{figure}

\section{Concluding remarks}

A realistic TCSM for fusion has been proposed, which is based on two spherical WS potentials
and the PSE method. This model describes the sp motion in a fusing system.
The adiabatic and diabatic molecular sp states for neutrons and protons
in the reaction $^{16}$O + $^{40}$Ca $\to$ $^{56}$Ni have been calculated.
A technique to compute the stationary diabatic states has been introduced,
which is based on the minimization of the strong radial nonadiabatic coupling
between the adiabatic states.
The only difference between the stationary diabatic sp basis and the adiabatic one is that
the diabatic basis minimizes the strong radial nonadiabatic coupling localized at some of the
avoided crossings between the adiabatic levels. Whether it is more convenient to use either
the adiabatic or the diabatic basis essentially depends on the bombarding energy \cite{Cassing1}.
It is important to add that the \textit{dynamical} radial nonadiabatic coupling is proportional
to the ion-ion radial velocity $\dot{R}$ which modulates the stationary (or static) couplings
calculated in the present work.

At very low incident energies (e.g., energies below the Coulomb barrier) one can
expect that the real sp motion is well described by the adiabatic representation. The dynamics
of the sp motion becomes even simpler when the dynamical nonadiabatic couplings may be neglected
and so the nucleons remain at the lowest energy levels. When the incident energy increases, the
dynamical couplings are more and more strengthened. In this case, nucleon transitions between the
adiabatic molecular sp levels occur. The adiabatic basis along with all nonadiabatic couplings
should then be used for understanding the real sp motion. If the incident energy reaches a certain
value (e.g., $1$ MeV/u above the Coulomb barrier for heavy symmetric systems \cite{Cassing1})
at which the dynamical nonadiabatic coupling is very strong
(probability of the nucleon transition is close to one), then the diabatic sp basis approximates
the real sp motion. At this energy regime, the use of the diabatic basis is certainly very useful
to simplify the dynamical calculation. It is worth mentioning that the diabatic sp motion is
expected to be realistic in the entrance phase of heavy-ion reactions \cite{Cassing1}.
The adiabatic and diabatic representations can be used as sp bases in
a coupled channels calculation within a molecular quantum-mechanical formulation of
heavy ion collisions \cite{Park,Imanishi}. Works regarding the application of the present TCSM
in understanding the formation mechanism of superheavy elements are in progress.


$\textit{Aknowledgements:}$ One of the authors (A.D-T) thanks M. H\"ol\ss,
P. Stevenson, I.J. Thompson,
S.N. Ershov, W. Greiner, C. Greiner and J.A. Maruhn for fruitful
discussions, and the Alexander von Humboldt Foundation for financial support.

 \appendix

\section{Actual radial nonadiabatic coupling matrix elements}
 \label{Radial_matrix_elements}

The matrix elements $<indx\ \Omega|\partial /\partial R |indx'\ \Omega>$ can be calculated
as follows including only the first four terms of (\ref{24b}) (for clarity $\Omega$ will be removed,
and new $\alpha$ and $\beta$ indexes will refer to $indx$ and $indx'$, respectively)

\begin{equation}
<\alpha|\frac{\partial}{\partial R}|\beta>=term1 + term2 + \cdots + term8, \label{appendx_1}
\end{equation}
where

\begin{equation}
term1=\sum_{\nu' = 1}^{N_1}\sum_{\nu = 1}^{N_1}
C_{1\nu' }^{*} < 1\nu'|\ G_0(E_{\alpha})G_0(E_{\beta})\ |1\nu>
\frac{\partial C_{1\nu }}{\partial R}, \label{appendx_term1}
\end{equation}

\begin{equation}
term2=\sum_{\nu' = 1}^{N_2}\sum_{\nu = 1}^{N_2}
C_{2\nu' }^{*} < 2\nu'|\ G_0(E_{\alpha})G_0(E_{\beta})\ |2\nu>
\frac{\partial C_{2\nu }}{\partial R}, \label{appendx_term2}
\end{equation}

\begin{equation}
term3=\sum_{\nu' = 1}^{N_1}\sum_{\nu = 1}^{N_1}
C_{1\nu' }^{*} < 1\nu'|\ G_0(E_{\alpha})\frac{\partial G_0(E_{\beta})}
{\partial R}\ |1\nu> C_{1\nu }, \label{appendx_term3}
\end{equation}

\begin{equation}
term4=\sum_{\nu' = 1}^{N_2}\sum_{\nu = 1}^{N_2}
C_{2\nu' }^{*} < 2\nu'|\ G_0(E_{\alpha})\frac{\partial G_0(E_{\beta})}
{\partial R}\ |2\nu> C_{2\nu }, \label{appendx_term4}
\end{equation}

\begin{equation}
term5=\sum_{\nu' = 1}^{N_1}\sum_{\nu = 1}^{N_2}
C_{1\nu' }^{*} < 1\nu'|\ G_0(E_{\alpha})G_0(E_{\beta})
e^{i \textbf{R} \hat{k}}\ |2\nu>
\frac{\partial C_{2\nu }}{\partial R}, \label{appendx_term5}
\end{equation}

\begin{equation}
term6=\sum_{\nu' = 1}^{N_2}\sum_{\nu = 1}^{N_1}
C_{2\nu' }^{*} < 2\nu'|\ G_0(E_{\alpha})G_0(E_{\beta})
e^{-i \textbf{R} \hat{k}}\ |1\nu>
\frac{\partial C_{1\nu }}{\partial R}, \label{appendx_term6}
\end{equation}

\begin{equation}
term7=\sum_{\nu' = 1}^{N_1}\sum_{\nu = 1}^{N_2}
C_{1\nu' }^{*} < 1\nu'|\ G_0(E_{\alpha})\frac{\partial G_0(E_{\beta})}
{\partial R}
e^{i \textbf{R} \hat{k}}\ |2\nu> C_{2\nu }, \label{appendx_term7}
\end{equation}

\begin{equation}
term8=\sum_{\nu' = 1}^{N_2}\sum_{\nu = 1}^{N_1}
C_{2\nu' }^{*} < 2\nu'|\ G_0(E_{\alpha})\frac{\partial G_0(E_{\beta})}
{\partial R}
e^{-i \textbf{R} \hat{k}}\ |1\nu> C_{1\nu }. \label{appendx_term8}
\end{equation}

The matrix elements contained in $term1$ and $term2$ (i.e.,
$< s\nu'|\ G_0(E_{\alpha})G_0(E_{\beta})\ |s\nu> $ with $s=1,2$)
are obtained using both the relation

\begin{equation}
G_0(E_{\alpha})G_0(E_{\beta})=\frac{1}{(E_{\alpha}-E_{\beta})}
[ G_0(E_{\alpha}) - G_0(E_{\beta}) ],\ E_{\alpha} \ne E_{\beta}
\label{relation1}
\end{equation}
and the matrix elements given by expression (\ref{eq_21a}).

The matrix elements of
$term3$ and $term4$ are calculated applying the relation

\begin{equation}
< s\nu'|\ G_0(E_{\alpha})\frac{\partial G_0(E_{\beta})}
{\partial R}\ |s\nu>  =
\frac{\partial E_{\beta}}{\partial R}
\frac{\partial}{\partial E_{\beta}}
< s\nu'|\ G_0(E_{\alpha})G_0(E_{\beta})\ |s\nu> .\nonumber \\
\label{relation2}
\end{equation}

The rest of the matrix elements involved in
$term5,\cdots,term8$ are obtained making use of (i) the
relation (\ref{relation1}), (ii) the matrix elements calculated
with expression (\ref{eq_21b}),
and (iii) the derivative rule used in the relation (\ref{relation2}).


\newpage

\begin{thebibliography}{00}




\bibitem{Holzer} P. Holzer, U. Mosel and W. Greiner, Nucl. Phys. A \bf138\rm (1969) 241.

\bibitem{Maruhn} J.A. Maruhn and W. Greiner, Z. Phys. \bf251\rm (1972) 431.

\bibitem{Park} W. Greiner, J.Y. Park and W. Scheid, in $\textit{Nuclear Molecules}$
(World Scientific, Singapore, 1994).

\bibitem{Mirea} M. Mirea, Phys. Rev. C \bf54\rm (1996) 302.

\bibitem{Radu} R.A. Gherghescu, Phys. Rev. C \bf67\rm (2003) 014309.

\bibitem{Pruess} K. Pruess and P. Lichtner, Nucl. Phys. A \bf291\rm (1977) 475.

\bibitem{Nuhn} G. Nuhn, W. Scheid and J.Y. Park, Phys. Rev. C \bf35\rm (1987) 2146.

\bibitem{Revai1} J. Revai, JINR, E4-9429, Dubna (1975).

\bibitem{Revai2} B. Gyarmati, A.T. Kruppa and J. Revai, Nucl. Phys. A \bf326\rm (1979) 119.

\bibitem{Gareev} F.A. Gareev, M.Ch. Gizzatkulov and J. Revai, Nucl. Phys. A \bf326\rm (1977) 512.

\bibitem{Milek} B. Milek and R. Reif, Phys. Lett. B \bf157\rm (1985) 134;
Nucl. Phys. A \bf458\rm (1986) 354.

\bibitem{Ershov} F.A. Gareev, S.N. Ershov, J. Revai, J. Bang and B.S. Nilsson,
Phys. Scripta \bf19\rm (1979) 509.

\bibitem{Gyarmati} B. Gyarmati and A.T. Kruppa, Nucl. Phys. A \bf378\rm (1982) 407;
B. Gyarmati, A.T. Kruppa, Z. Papp and G. Wolf, Nucl. Phys. A \bf417\rm (1984) 393.

\bibitem{Fonseca} A.C. Fonseca, J. Revai and A. Matveenko, Nucl. Phys. A \bf326\rm (1979) 182;
J. Revai and A. Matveenko, Nucl. Phys. A \bf339\rm (1980) 448.

\bibitem{Delos} J.B. Delos and W.R. Thorson, J. Chem. Phys. \bf70\rm (1979) 1774;
Rev. Mod. Phys. \bf53\rm (1981) 287.

\bibitem{Cassing1} W. Cassing and W. N\"orenberg, Nucl. Phys. A \bf433\rm (1985) 467.

\bibitem{Alexis1} A. Diaz-Torres, Phys. Rev. C \bf69\rm (2004) 021603(R).

\bibitem{Alexis3} A. Diaz-Torres, Phys. Lett. B \bf594\rm (2004) 69.

\bibitem{Ring} P. Ring and P. Schuck, in $\textit{The Nuclear Many-Body Problem}$
(Springer-Verlag, New York Inc., 1980) p. 41.

\bibitem{Varshalovich} D.A. Varshalovich, A.N. Moskalev and V.K. Khersonskii,
in $\textit{Quantum Theory of Angular Momentum}$
(World Scientific, Singapore, 1989).

\bibitem{Neumann} J. von Neumann and E. Wigner, Z. Physik \bf30\rm (1929) 467.

\bibitem{Nikitin} E.E. Nikitin and S.Ya. Kmanskii, in $\textit{Theory of
Slow Atomic Collisions}$ (Springer-Verlag, Berlin, 1984) p.74.

\bibitem{Lukasiak1} A. Lukasiak, W. Cassing and W. N\"orenberg, Nucl. Phys. A \bf426\rm
(1984) 181.

\bibitem{Alexis2} A. Diaz-Torres, N.V. Antonenko and W. Scheid, Nucl. Phys. A \bf652\rm
(1999) 61.

\bibitem{Diabbasis} A. Troisi and G. Orlandi, J. Chem. Phys. \bf118\rm (2003) 5356,
and references therein.

\bibitem{Soloviev} V.G. Soloviev, in $\textit{Theory of
Atomic Nuclei (Nuclear Models)}$ (Energoizdat, Moscow, 1981) p.67.

\bibitem{Irving} J.M. Irvine, in $\textit{Nuclear Structure Theory}$
(Pergamon Press, Oxford, 1972) p.313.

\bibitem{Terlecki} G. Terlecki, W. Scheid, H.J. Fink and W. Greiner,
Phys. Rev. C \bf18\rm (1978) 265.

\bibitem{Imanishi} B. Imanishi and W. von Oertzen, Phys. Rep. \bf155\rm (1987) 29.

\end{thebibliography}
\end{document}